\documentclass[prl,aps,twocolumn, floatfix, superscriptaddress,showpacs]{revtex4}
\usepackage{amsmath,bm,graphicx}
\usepackage{amssymb}
\usepackage{amsfonts}
\usepackage{color}
\usepackage{ulem}
\usepackage{natbib}

\begin{document}

\title{Bistability, oscillations and bidirectional motion of ensemble of hydrodynamically-coupled molecular motors}

\author{P. Malgaretti}
\email[Corresponding Author : ]{malgaretti@is.mpg.de }
\affiliation{Max-Planck-Institut f\"{u}r Intelligente Systeme, Heisenbergstr. 3, D-70569
Stuttgart, Germany}
\affiliation{IV. Institut f\"ur Theoretische Physik, Universit\"{a}t Stuttgart,
Pfaffenwaldring 57, D-70569 Stuttgart, Germany}
\author{I. Pagonabarraga}
\affiliation{Departament de Fisica de la Mat\`eria Condensada, Facultat de Fisica, Universitat de Barcelona, Carre Mart\'i i Franques 1, Barcelona 08028, Spain}
\affiliation{UBICS, Institute of Complex Systems,  Universitat de Barcelona, Barcelona, Spain}
\affiliation{CECAM, Centre Europ\'een de Calcul Atomique et Mol\'eculaire, \'Ecole Polytechnique F\'ed\'erale de Lasuanne, Batochime, Avenue Forel 2, 1015 Lausanne}
\author{J.-F. Joanny}
\affiliation{Physicochiemie Curie (Institut Curie/CNRS-UMR168/UPMC), Institut Curie, Centre de Recherche PSL Reseach University, 26 rue d'Ulm 75248 Paris Cedex 05, France}
\affiliation{ESPCI 10 rue Vauquelin 75005 Paris, France}

\begin{abstract}

We analyze  the collective behavior of hydrodynamically coupled molecular motors. We show that the local fluxes induced by motors displacement can induce the experimentally observed bidirectional motion of cargoes and vesicles. 
By means of a mean--field approach we show that  sustained oscillations as well as bistable collective motor motion arise even for very large collection of motors, when thermal noise is irrelevant.  The analysis clarifies the physical mechanisms responsible for such dynamics by identifying the relevant coupling parameter and its dependence on the geometry of the hydrodynamic coupling as well as on system size. We quantify the phase diagram for the different phases that characterize the collective motion of hydrodynamically coupled motors and show that sustained oscillations can be reached for biologically relevant parameters, hence demonstrating the relevance of hydrodynamic interactions in intracellular transport.

\end{abstract}
%
\pacs{87.16.Nn,87.16.Wd,47.63.-b}
\keywords{===}

\maketitle

The transport of macromolecules, vesicles and organelles inside cells relies on the active motion of molecular motors along biofilaments~\cite{Howard}. When motors  pull on  organelles or vesicles, the  fluid flow that they  induce provides motor-motor hydrodynamic interactions (HI) hence providing an additional mechanism for molecular motors coupling~\cite{Malgaretti} that, possibly, is responsible of the experimentally observed enhanced velocity of motors pulling on fluid vesicles~\cite{Nelson2014}. 
In particular, the motion of these cargoes has been observed to be mono- and bi- directional, the latter relying on the presence of motors pulling the cargoes in opposite directions~\cite{Ma,Warshaw,Kunwar2011,badoual,Nebenfuhr01121999,Hendricks,goldman2010}. 
While previous studies have focused preferentially on rigidly coupled motors~\cite{Julicher1995,Julicher1997,Lipowsky2010, Guerin2011}, in this letter we show that the bidirectional motion observed in experiments can be induced and controlled via the HI induced by motors active displacement. Exploiting a mean-field approach we identify the key parameters controlling the onset of the bidirectional motion and we show the relevance of HI for biologically relevant scenarios.
%
%

While, a few motors at the cargo tips can  pull against the cargo drag, the rest of the motors can move along the cargo. Their net motion is responsible for the onset of HI. 
We model the molecular motor  dynamics exploiting the two-state-model~\cite{Julicher1995} that regards molecular motors as particles with two internal states. 
In order to account for motors pulling in opposite directions, we model the two families of motors as a single family of \textit{effective} motors that in the ``bound'' state experience a symmetric periodic force $f(x)=-\partial_x V=f_0\cos (2\pi x/L)=f_0\tilde{f}(x)$ of period $L$~\cite{Guerin2011}. With rate $\omega_{off}(x)$ motors  jump to a ``weakly bound'' state, in which they diffuse freely. Motors bind with rate $\omega_{on}(x)$.
We describe the system in terms of  densities of bound ($\rho(x)$) and weakly bound ($\sigma(x)$) motors. The total density, $\rho(x)+\sigma(x)$, cannot exceed the maximal  filament occupation prescribed by excluded volume, and in order to keep analytical insight, we    assume that the motors  remain in a  dilute regime. The motion of molecular motors occurs in the low Reynolds regime and it generates a fluid flow whose magnitude reads
\begin{equation}
\gamma v(x)= f(x) +\int f(y)\rho(y)W(x,y)dy 
\label{vel} 
\end{equation}
where $W(x,y)$ is the dimensionless Oseen tensor accounting for HI~\footnote{The Oseen  approximation is valid for  dilute systems and constitutes a lower limit for  denser systems (see Suppl. Mat.)} and $\gamma$ is the single motor drag coefficient. According to Eq.~(\ref{vel}),  the induced fluid velocity is characterized by the local flow due to the force that the filament exerts on the molecular motor (first term  in rhs of Eq.~\ref{vel}) and the collective hydrodynamic flow induced by the rest of the molecular motors at the position of the reference motor. This separation between local and collective induced flows is characteristic of softly interacting motors, and it is absent for rigidly coupled motors~\cite{Guerin2011}. Since the HI extends over distances large compared  to the spatial variation of the motor-filament interaction, for a large system size $\Lambda$, $\Lambda \gg L$, we assume a mean field HI
\begin{equation}
\begin{aligned}
 \int_\Lambda f(y)\rho(y)W(x,y)dy & \simeq  k\frac{L}{\Lambda}\int_{\Lambda} f(y)\rho(y)dy+O(1)\\
k & =  \frac{1}{L}\int_{\Lambda} W(x,y)dy,
\label{vel-approx}
\end{aligned}
\end{equation}
which  holds when $f(x)=f_0$ and $\rho(x)=\rho_0$, and becomes exact for an infinite system, $\Lambda \rightarrow \infty$, if  the motor density follows the spatial dependence imposed by the  motor/filament periodic force. In this regime, reasonable when boundary effects can be disregarded, the  spatial dependence of the hydrodynamic long-range coupling, $W(x,y)$,  experienced by the motors becomes negligible and it 
can be described by an effective  dimensionless parameter, $k$. For example, for motors of linear size $R$ moving in a three-dimensional environment, $W(x,y)=\frac{3R}{2}\frac{1}{|x-y|}$, leading to
\begin{equation}
k_{3D}= \int_{2R}^{\Lambda/2} \frac{3R}{2}\frac{dr}{r}=\frac{3}{2}\frac{R}{L}\ln(\Lambda/4R).
\label{k_3D}
\end{equation}
which grows logarithmically with system size.
Alternatively, when motors are pulling on a membrane-coated cargo, such as organelles or vesicles, their tails are linked to molecules embedded in the fluid-like membrane (characterized by a $2D$ viscosity $\eta_{2D,mem}$) quite more viscous than the cytoplasm (characterized by a $3D$ viscosity $\eta_{3D,cyt}$). When the intrinsic dynamics of motor tails does not affect motors dynamics, we can identify the dynamics of the motors with that of the tracers and therefore we can calculate the $2D$ HI. In this regime, we expect the $2D$ HI to dominate  the three-dimensional flows induced in the cytoplasm for motor-motor separations smaller than $l=\eta_{2D,mem}/\eta_{3D,cyt}$. 
It is then reasonable to assume  $k=k_{3D}+k_{2D}$ with
\begin{equation}
\frac{k_{2D}}{2}= \int_{2R}^{M}\ln\frac{l}{r}dr=\frac{M}{L}\left(1+\ln\frac{l}{M}\right)-2\frac{R}{L}\left(1+\ln\frac{l}{2R}\right) 
\label{k_2D}
\end{equation}
where $2\ln\frac{l}{r}$ is the dimensionless Oseen tensor in $2D$ and $M=\min\left(\frac{l}{e},\frac{\Lambda}{2}\right)$. Eqs.~(\ref{k_3D}),(\ref{k_2D}) capture the diverging nature of HI for increasing system sizes $\Lambda$. Therefore, for large system sizes, $\Lambda/L\gg 1$, we can neglect the $O(1)$ terms in Eq.~(\ref{vel-approx}) and Eq.~(\ref{vel}) reduces to
\begin{equation}
v(x)=\frac{f_0}{\gamma}\left(\tilde{f}(x)+k\left\langle \tilde{f}(x)\tilde\rho(x)\right\rangle _{x}\right)
\end{equation}
where we have introduced the dimensionless density $\tilde\rho(x)=L \rho(x)$  (and similarly we introduce $\tilde\sigma(x)=L \sigma(x)$).

Accordingly, we can write, in dimensionless units, the evolution of the mean bound, 
$\hat\rho(x)=\left\langle {\tilde\rho}(x)\right\rangle_{\tilde\rho,\tilde\sigma}$, and weakly bound, $\hat\sigma(x)=\left\langle {\tilde\sigma}(x)\right\rangle_{\tilde\rho,\tilde\sigma}$, motor densities~\footnote{Eqs.~(\ref{fokker-planck}) are obtained via a mean--field approximation (See Suppl. Mat.).
}
\begin{eqnarray}\label{fokker-planck}
\dot{\hat\rho}(x) &=&-\partial_{x}\lambda \hat\rho(x)\left[\tilde{f}(x)+k\left\langle \tilde{f}(x) \hat\rho(x)\right\rangle _{x}\right]+\nonumber\\
&&-\tilde\omega_{off}(x)\hat\rho(x)+\tilde\omega_{on}(x)\hat\sigma(x)\nonumber\\
\dot{\hat\sigma}(x) &=&-\partial_{x}\lambda\hat\sigma(x)k\left\langle \tilde{f}(x) \hat\rho(x)\right\rangle_{x}+\\ &&+\tilde\omega_{off}(x) \hat\rho(x)-\tilde\omega_{on}(x) \hat\sigma(x)+\nonumber\\
&&+k_{\mbox{on}}\left( c_{\infty}-\hat{\rho}(x)-\hat{\sigma}(x)\right)-k_{\mbox{off}} \hat{\sigma}(x)\nonumber
\end{eqnarray}
where $\left\langle \psi(x)\right\rangle_{\rho}=\int \psi P[\rho(x)] dx$ stands for the average of $\psi$ over the probability distribution of density profiles $\rho(x)$ while $\left\langle \psi(x)\right\rangle_{x}=(1/L) \int_0^L \psi(x) dx$ corresponds to the spatial average of $\psi$ over the filament period (see Suppl. Mat).  The first term in the rhs of Eqs.~(\ref{fokker-planck}) describes motor advection according to the  local velocity the motors are exposed to, the second term describes the  motor binding kinetics while the third term in the evolution equation for weakly bound motors describes the fact that motors can  detach from the filament with rate $k_{\mbox{off}}$ and bind with rate $k_{\mbox{on}}$, proportional to the available space $c_{\infty}-\hat{\rho}(x)-\hat{\sigma}(x)$ where $c_{\infty}$ is the motor concentration in bulk.~\footnote{Since bound and weakly bound species correspond to different states of the same motor, their sum cannot exceed a maximum density, $\zeta_{max}$, corresponding to complete local occupancy. Accordingly, for non-dilute regimes, a global factor on the rhs of Eq.~(\ref{fokker-planck}) must be included to ensure such a constraint.}.


The motor flux in Eqs.~(\ref{fokker-planck}) is proportional to $\lambda=f_0/\bar{\omega} L \gamma$, the ratio of the typical time a motor needs to slide down the potential, $L \gamma/f_0$, and the characteristic inverse hopping time $\bar{\omega}=\langle \omega_{on}(x)+\omega_{off}(x)\rangle_x/2$.
The bound and weakly bound rate densities depend on the filament structure. We consider the simple, periodic form
\begin{equation}
 \tilde{\omega}_{on/off}=\max(\Delta \omega_{on/off}\mp\delta \omega_{on/off}\sin(2\pi x/L),0)
 \label{rates}
\end{equation}
which interpolates between  regimes where the hopping rates are essentially homogeneous along the filament, $\Delta \omega_{on/off}\gg\delta \omega_{on/off}$,  or only take place at  localized regions  on the filament, $\Delta \omega_{on/off}\ll\delta \omega_{on/off}$. The latter, together with an opposite sign between the bound and weakly bound rates accounts for the fact that binding and unbinding processes are  localized at different  regions along the filament. Normally,  unbinding rates are  more spatially localized than bounding rates, implying $\delta \omega_{off}/\Delta \omega_{off} > \delta \omega_{on}/\Delta \omega_{on}$.

For fast bulk kinetics $k_{\mbox{on,off}}\rightarrow\infty$,  the concentration of weakly bound motors, $\hat\sigma$ is homogeneous along the filament. In this regime, Eqs.~(\ref{fokker-planck}) decouple and the collective behavior of the motors is controlled by the  evolution of $\hat\rho(x)$, 
\begin{align}
 \dot{\hat\rho}(x)&=-\lambda\partial_{x}\hat\rho(x) \left[\tilde f(x)+k\left\langle  \tilde f(x) \hat\rho(x) \right\rangle _{x}\right]+\nonumber\\
-&\tilde{\omega}_{off}(x)\hat\rho(x)+\frac{k_{\mbox{on}}}{k_{\mbox{off}}+k_{\mbox{on}}}\tilde{\omega}_{on}(x)\left(c_\infty-\hat\rho(x)\right)
\label{fokker-planck-1field}
\end{align}
in terms of  rescaled  binding rates that keep the same functional dependence as in Eq.~(\ref{rates}).
\begin{figure}
 \includegraphics[scale=0.22]{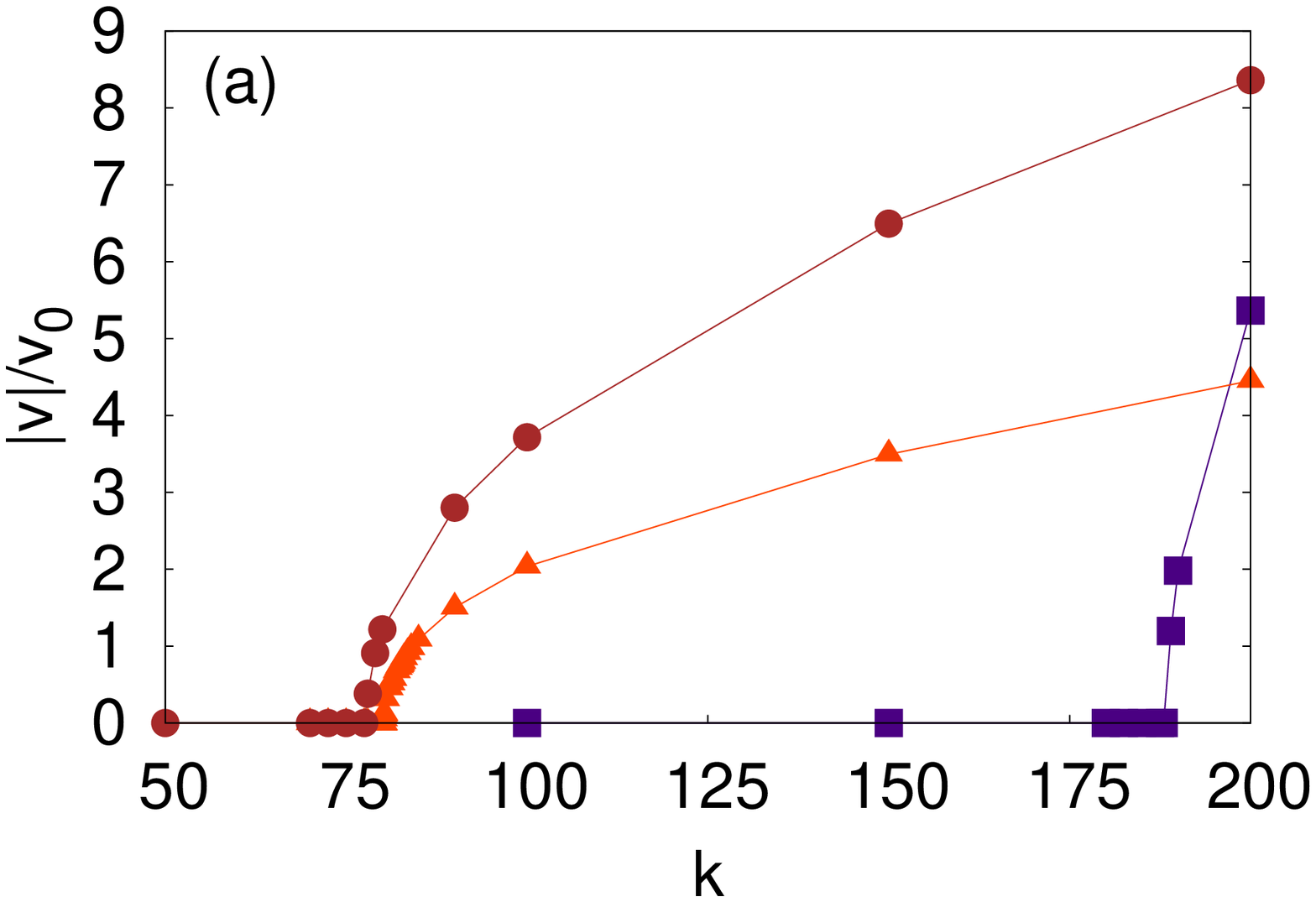}\,\includegraphics[scale=0.22]{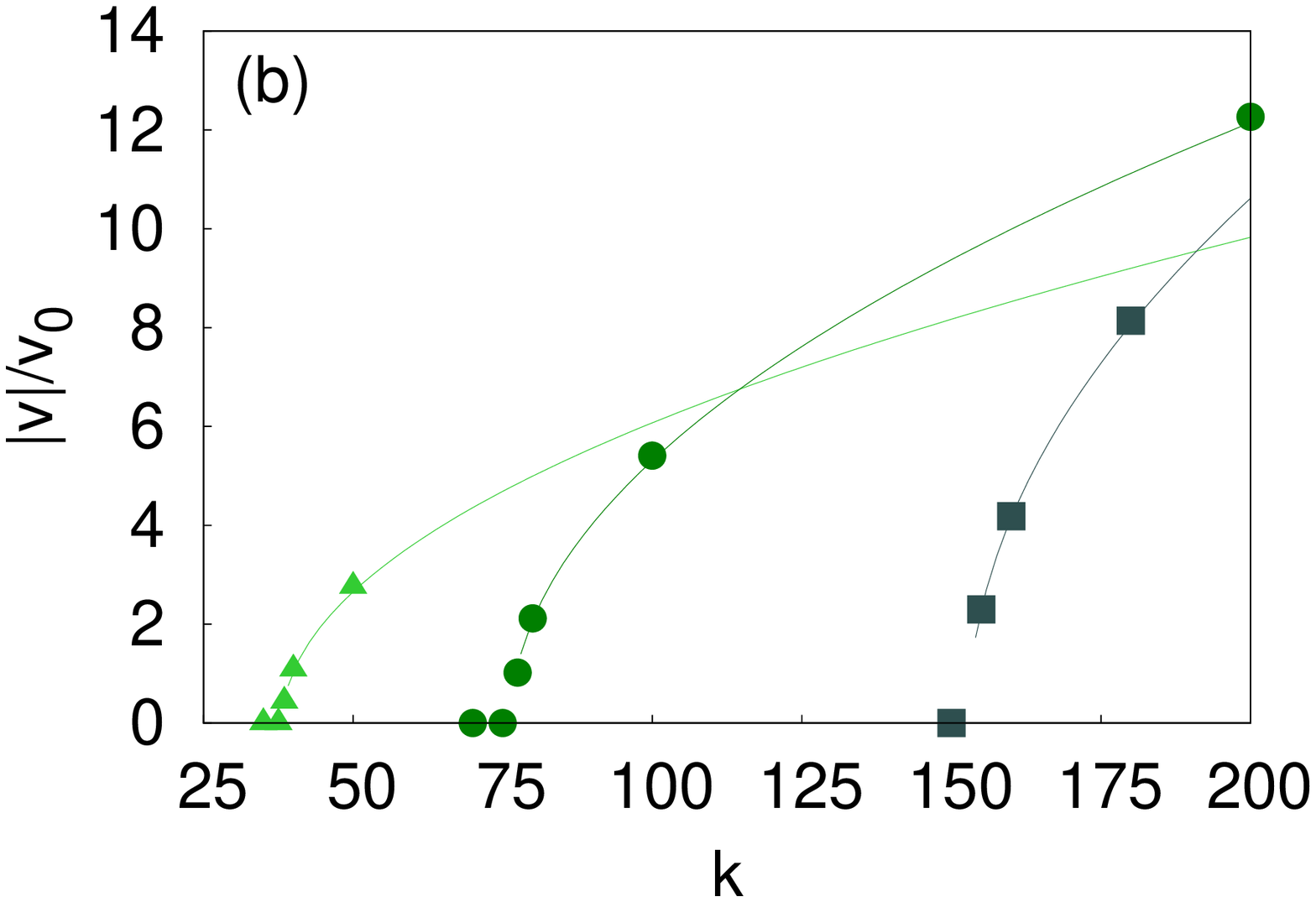}
 \caption{Average molecular motor velocity normalized by the single bound motor velocity, $v_0=f_0/\gamma$, as a function of the dimensionless coupling $k$, as obtained from Eq.~\ref{fokker-planck-1field}. Hydrodynamically (panel a) and rigidly  (panel b) motor coupling, both characterized by $\Delta\omega_{on}=\Delta\omega_{off}=1$ and $\delta\omega_{on}=\delta\omega_{off}=1/2$ and $c_\infty=1$.   $\lambda=0.015,0.05,0.1$, (square, circle, triangle) and  $\lambda=0.017,0.033,0.066$ (square, circle, triangle) for panel (a) and (b) respectively.}
 \label{fig1}
\end{figure}
We have numerically solved Eq.~(\ref{fokker-planck-1field}) using a Lax-Wendroff scheme with periodic boundary conditions applied at the ends of a period of the ratchet potential. Fig.~\ref{fig1}.(a) shows that the configuration where motors  do not have a net velocity, stable for weak HI, becomes unstable  at a critical coupling, $k_o$, above which  a net motor current, that breaks left/right  symmetry, develops;  a similar scenario has been described for  rigidly coupled motors~\cite{Guerin2011}. Despite the apparent similarity between the emergence of net motion for soft and rigidly coupled motors,  the different underlying physical mechanisms responsible for symmetry braking lead to significant differences in collective motor dynamics. 
While hydrodynamically coupled motors are characterized by a non-monotonous dependence of $k_o$ on $\lambda$, see Fig.~\ref{fig1}.(a), the opposite holds true for rigidly coupled motors, Fig.~\ref{fig1}.(b)~\footnote{The rigid coupling regime is obtained by solving Eq.~(\ref{fokker-planck-1field}) disregarding  the local contribution, $\tilde f(x)$.}. 
A linear stability analysis around the quiescent  state  shows
\begin{equation}
 k>\frac{4}{\pi \lambda\left(\frac{\delta \omega}{\Delta \omega}-\pi\lambda\right)c_\infty}
\label{nec_cond}
\end{equation}
as the sufficient condition~\footnote{Eq.~(\ref{nec_cond}) constitutes only a sufficient and not necessary condition since it has been obtained  imposing that the dynamic system of linearized perturbations around the quiescent state has at least three real solutions without constraining their signs (see Suppl. Mat.)}. for symmetry breaking and onset of net motor currents.
Since $k\ge0$, Eq.~(\ref{nec_cond}) identifies an interval $\lambda \epsilon [0;\lambda_{max}]$ for which symmetry breaking occurs, with $\lambda_{max}= \frac{1}{\pi}\frac{\delta \omega}{\Delta \omega}$. According to Eq.~(\ref{nec_cond}), $k$ diverges for both $\lambda=0,\lambda=\lambda_{max}$ leading to the existence of a minimum value $k=k_c$ at $\lambda=\lambda_c=\frac{1}{2}\lambda_{max}$. For the parameters used in Fig.~\ref{fig1}.(a) the stability analysis predicts $\lambda_c\sim 8\cdot 10^{-2}$ and $k_c \lesssim 70$ in good agreement with numerical results.  For rigidly bound motors, the second term in the denominator of Eq.~(\ref{nec_cond}) disappears, leading to an inverse proportionality between the coupling constant $k$ and the dimensionless forcing $\lambda$, consistent with the results shown in Fig.~\ref{fig1}.(b). This difference has significant implications. For example,while  for $\lambda>\lambda_c$ a reduction in $\lambda$ will favor the onset of net fluxes, when    $\lambda < \lambda_c$ decreasing $\lambda$ will hinder, or even prevent, the development of a net motor flux.

\begin{figure}
 \includegraphics[scale=.22]{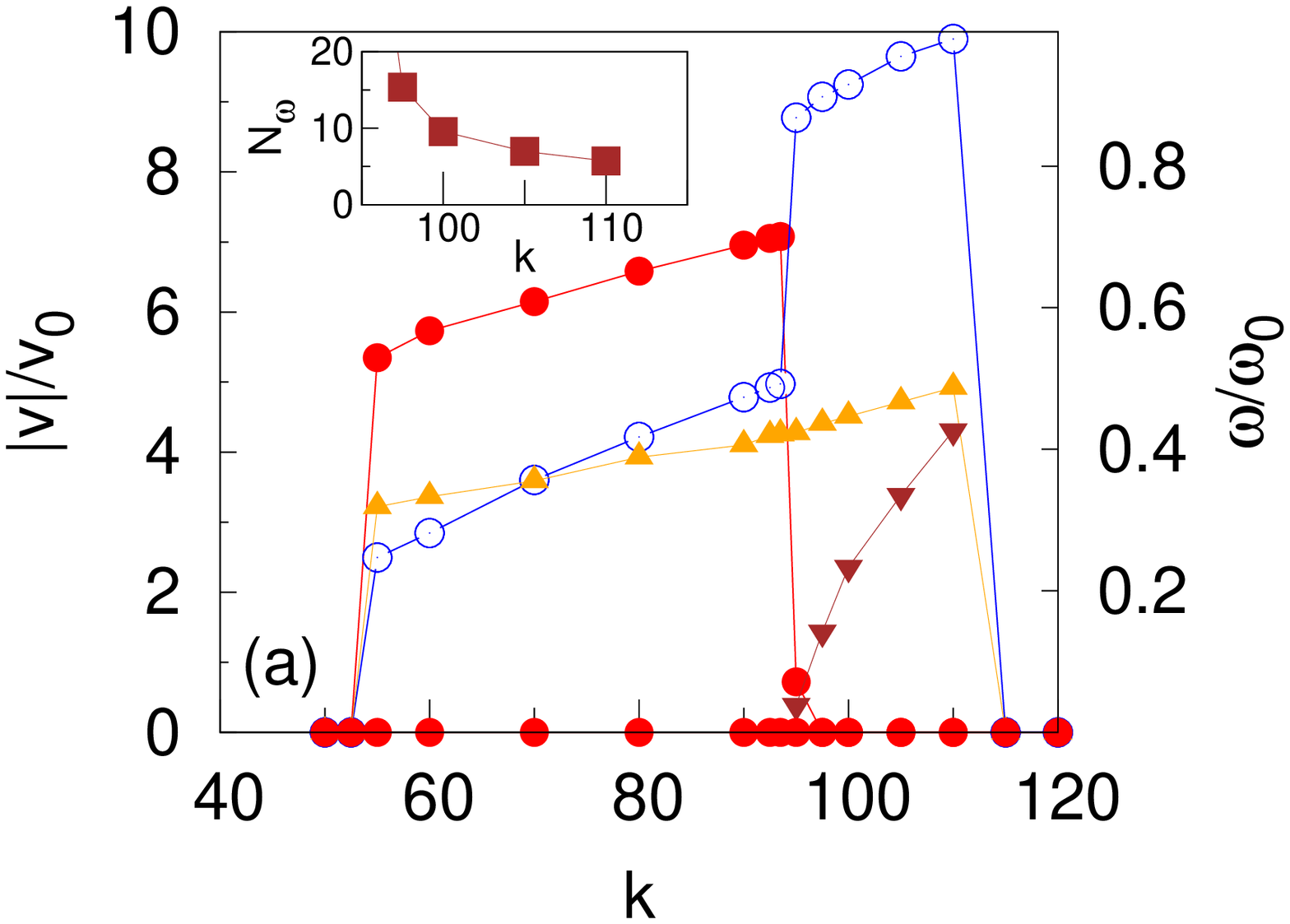}
 \includegraphics[scale=.22]{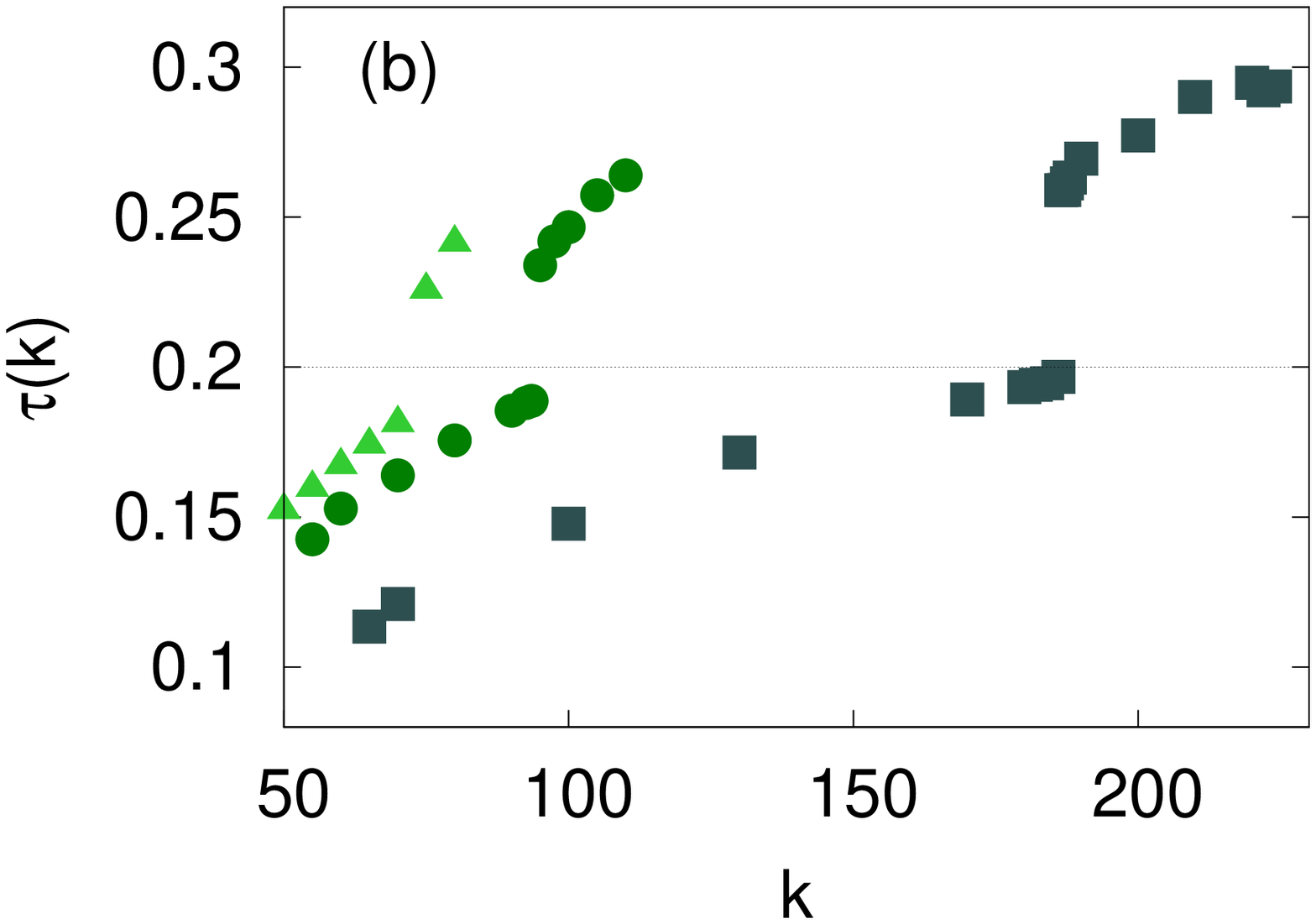}
\caption{(a): Average velocity (filled circles), velocity variance (open circles) (both normalized by the bound single motor velocity, $v_0=f_0/\gamma$), and sustained oscillations frequency, $\omega_v$ (upward triangles) and frequency of inversion in the bistable regime $\Omega_v$ (downward triangles) (both normalized by  the hopping rate peak value $\omega_0$) upon variation of $k$. $\Omega_v$ has been magnified by a factor of $10$ for sake of clearness. Hopping rates are characterized by $\Delta\omega_{on,off}=-1$, $\delta\omega_{on,off}=2$. Inset: number of oscillation between subsequent velocity switches, $N_\omega=2\omega_v/\Omega_v$, in the bistable regime. (b): dimensionless time $\tau(k)=2 v(k)/\omega_0\epsilon$ governing the stability of the sustained oscillation (see text) as a function of the dimensionless coupling $k$ for $\lambda=f_0/\bar{\omega}
L \gamma=2\cdot10^{-2},3.3\cdot10^{-2},4\cdot10^{-2}$ (squares, circles, triangles). Points below the dashed lines are for  motors undergoing sustained oscillation whereas above th dashed line for motors in the bistable regime.}
\label{fig2}
\end{figure}

In the opposite regime, when the exchange of molecular motors with the bulk is negligible, $k_{\mbox{on,off}}=0$, the total number of  motors moving along a filament is conserved.
When the evolution of bound, $\hat\rho(x)$, and weakly bound, $\hat\sigma(x)$, motors are coupled, Eqs.~(\ref{fokker-planck}) must be solved consistently. In this regime motors tend to accumulate spatially, leading, in some cases,   to large local motor densities. It is known that conservation of the overall number of motors moving along a filament promotes cluster~\cite{Malgaretti} and shock wave formation~\cite{Aghababaie}. The morphological details of these structures are sensitive to  excluded volume and short range interactions. However, for binding and unbinding rates sharply peaked  at the potential extrema, $\Delta\omega_{on,off}<\delta\omega_{on,off}$, the development of regions of high molecular motor density only occurs for large coupling parameters. Hence, we can address the instability of the homogeneous, quiescent molecular motor profile  avoiding the development of shock waves. We observe that this quiescent  configuration destabilizes above a threshold coupling parameter, $k_1$, characterized by a Hopf bifurcation, as shown in  Fig~\ref{fig2}.(a) (for $k_1=55$). Above $k_1$ the stable state is characterized by  a non-zero mean velocity and an oscillation of frequency, $\omega_v$, as shown in Fig.~\ref{fig2}.(a)~\footnote{For peaked hopping rates, the relevant time scale determining the oscillations is controlled by the hopping rates, $\omega_0$ rather than by their average values, $\bar{\omega}$}. Motor velocity oscillations emerge as a result of  the periodic change in the density of the bound motors. While moving under the action of the driving potential, the fluid flow generated by bound motors  advects  weakly bound motors along. After reaching the bottom of the potential, bound motors cease to move and jump   to the diffusive state with rate $\omega_{off}$. This leads to an increase of  diffusive motors, hence inducing an overall decrease in the average motor velocity, which relies in the small fraction of bound motors still displacing.  Once diffusing motors reach the hopping region, they bind strongly to the filament at a rate  $\omega_{on}$, starting a new cycle. As a result of this alternate, correlated motor exchange, both strongly, $\hat{\rho}(x)$, and weakly bound, $\hat{\sigma}(x)$, motor densities develop traveling waves.

As we increase $k/\lambda$ a second bifurcation is observed,  $k_2$, above which molecular motors exhibit bistability. In this regime the motor density increases gradually  where the  motor states can be switched, leading eventually to  an overall motor density that  exceeds the maximum occupancy. 
In order to explore this regime we then  redistribute uniformly the motor excess 
\footnote{When exceeding the maximum value, $\hat\rho_{max}$, the excess of motors is equally redistributed along the period.}.

For the system shown in Fig.~\ref{fig2}, above the threshold value $k_2=95$ the motor velocity still oscillates with frequency $\omega_v$ around a non vanishing mean velocity, $v$. However, in contrast to the regime $k_1< k<k_2$, where the sign of $v$ is fixed, for $k>k_2$ the sign of $v$ changes with frequency $\Omega_v$. For time scales larger than $\Omega_v^{-1}$, the average motor  velocity  vanishes and a bistable behavior emerges (see Fig.1 in Suppl. Mat.), analogous to the one experimentally observed~ \cite{Ma,Warshaw,Kunwar2011,badoual,Nebenfuhr01121999,Hendricks,goldman2010,Guerin-review}. 
This second transition is captured by the dimensionless time, $\tau(k)=2 v(k)/\omega_0\epsilon$, defined as the ratio between the characteristic hopping time $\frac{1}{2}\omega_0^{-1}$\footnote{The prefactor $\frac{1}{2}$ accounts for the  hopping rate variable shape.}, and the time a particle spends in the region in which the hopping rate is non vanishing, $\epsilon$, being pushed at a speed $v(k)$~\footnote{$v(k)$ is defined as the average velocity, $\left\langle v(k) \right\rangle_{x,\tilde \rho, \tilde \sigma}$, for $k<k_2$ whereas $v(k)=\sqrt{\left\langle v^2(k)\right\rangle_{x,\tilde \rho, \tilde \sigma}}$ for $k>k_2$, i.e. when $\left\langle v(k) \right\rangle_{x,\tilde \rho, \tilde \sigma}=0$.}. As shown in Fig.~\ref{fig2}.(b), when $\tau(k)\lesssim 0.2$ the time needed to jump between the two states is much smaller than the time that a particle spends in crossing the hopping region, therefore the majority of the motors rebind and the system undergoes sustained oscillations. On the contrary, when $\tau(k) \gtrsim 0.2$ part of the motors in the weakly bound state cannot jump back to the bound state and do not contribute to the next cycle. The loss of active motors affects subsequent oscillations. These effects sum up until the system switches the direction of the average velocity on time scales of the order of $\Omega_v$. The number of oscillations between two subsequent switching events,  $N_{\omega}=2\omega_v/\Omega_v$, decreases for 
increasing $k$, as shown in the inset of Fig.~\ref{fig2}.(a). For increasing $k$ the two time scales approach, $N_{\omega}\rightarrow 1$, and the bistability disappears; subsequently motors remain in a quiescent state.

The bistable state  observed is typical of soft, hydrodynamically coupled motors for which the local density of both bound and weakly bound motors can adjust dynamically. 
This feature is absent both if  we disregard weakly bound motor dynamics (see Fig.~\ref{fig1}), or for rigidly coupled motors~\cite{Guerin2011},  when motor density rearrangements are suppressed. Bistability can be recovered  for  rigidly coupled motors~\cite{Guerin2011}, also when weakly bound motors are in contact with a reservoir (data not shown), only if the hopping dynamics is noisy. However, this mechanism vanishes for large system sizes, for which the noise becomes negligible. On the contrary, for hydrodynamically coupled motors  bistability arises by increasing the system size, encoded in $k$, and persists for a finite range of values of $k$ as shown in Fig.~\ref{fig2}.(a), regardless of system size.
\begin{figure}
 \includegraphics[scale=0.22]{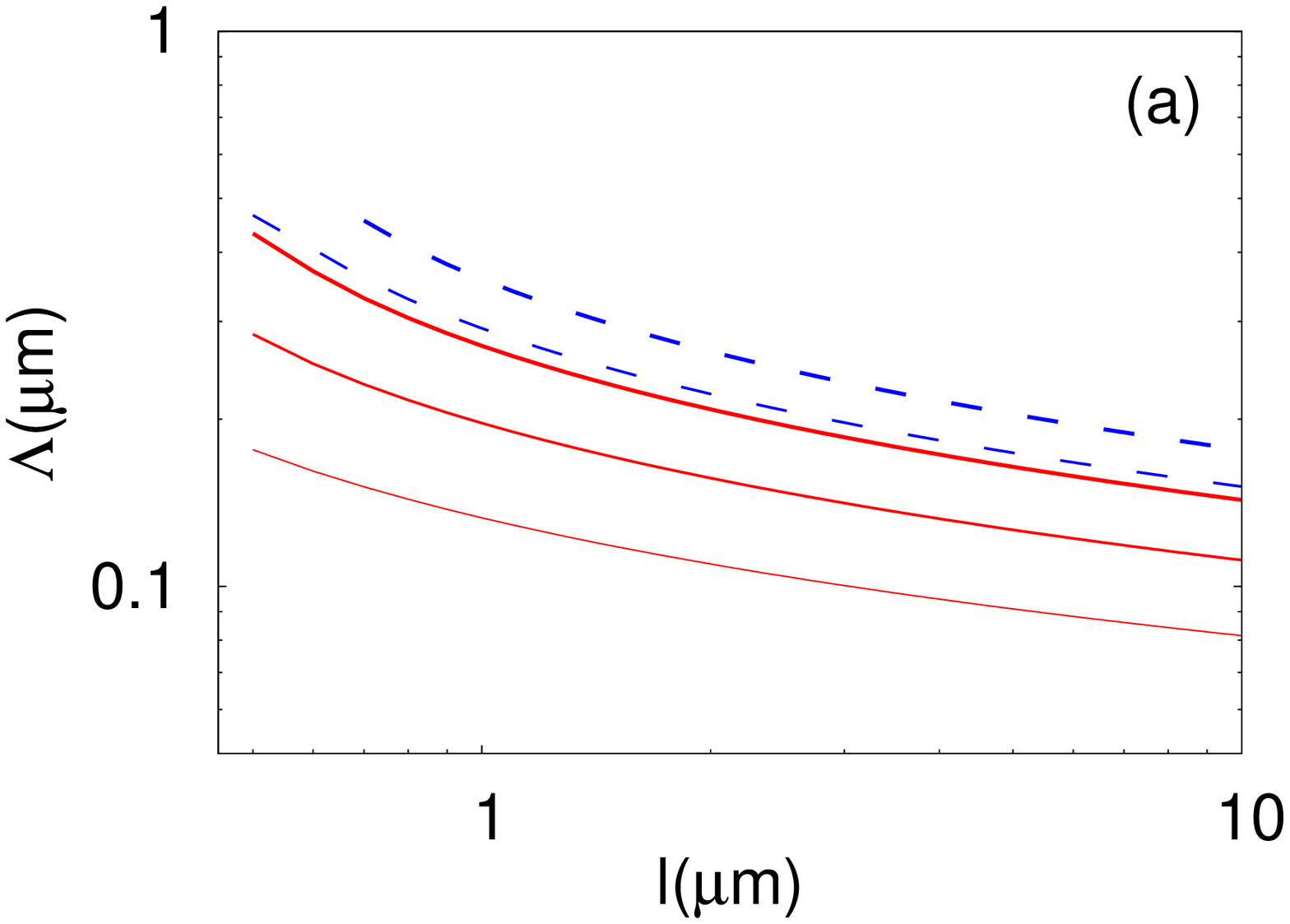}
 \includegraphics[scale=0.22]{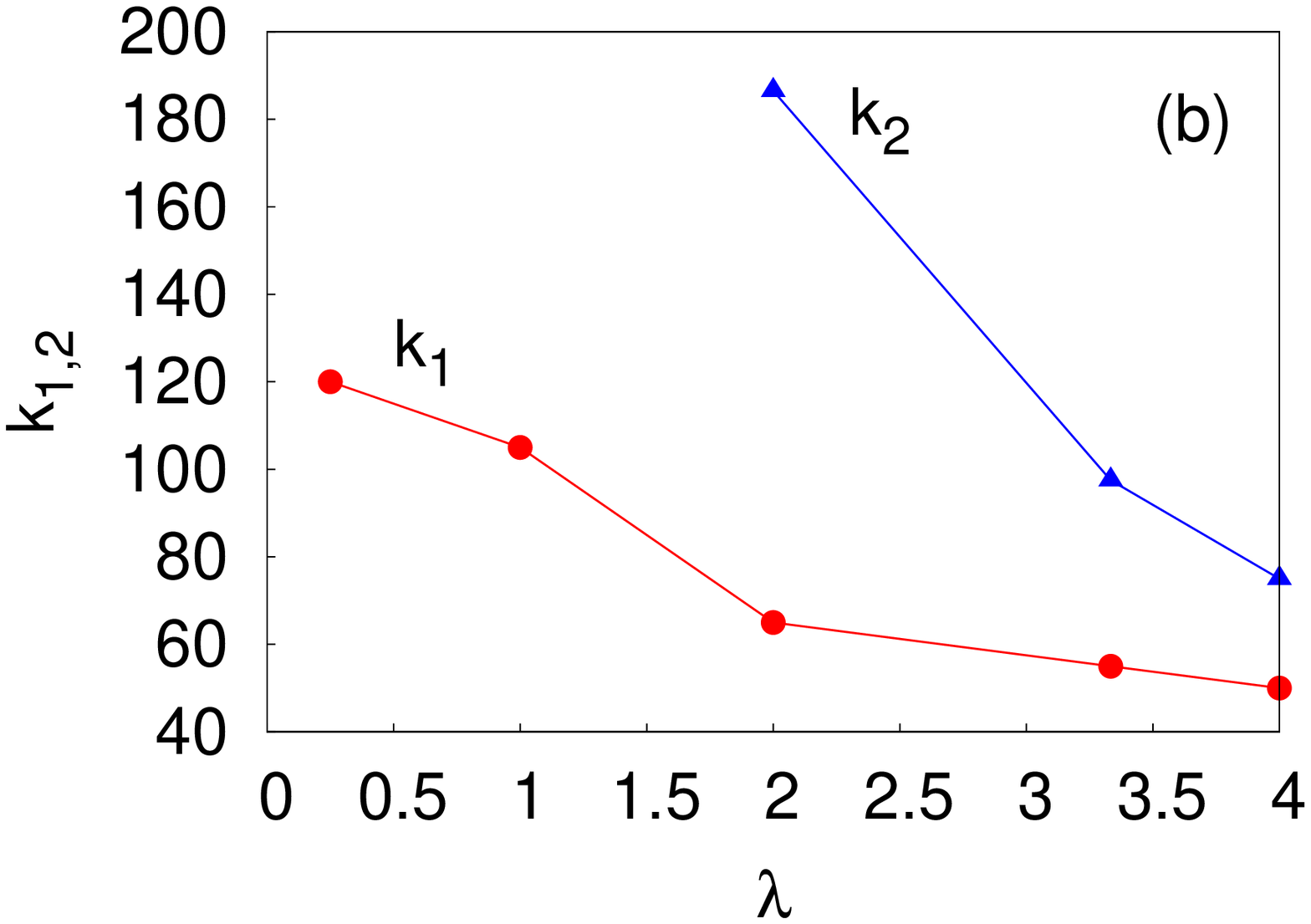}
\caption{(a): minimum system size, $\Lambda$, calculated from Eq.~(\ref{k_2D}) for symmetry breaking (solid lines) and for bistable onset (dashed lines) as a function of $l=\eta_{2D,mem}/\eta_{3D,cyt}$ for motor pulling on membranes. The membrane-embedded tracer size, $R$, is of the order of the membrane thickness $\sim 4 nm$ leading to  $R\sim 1/2 L$. Thicker lines stands for larger values of $k$: $k=50,70,90$ (solid lines) and $k=95,110$ (dashed lines) respectively. (b): values of the coupling parameters, $k_1,k_2$ at which the two bifurcations occur as a function of the motor properties encoded in $\lambda=f_0/(\bar{\omega} L \gamma)$.}
\label{symmetry_breaking}
\end{figure}

The collective phases identified for hydrodynamically coupled motors are controlled  by the coupling parameter $k$ that, according to Eqs.~(\ref{k_3D}),(\ref{k_2D}), depends on the system size $\Lambda$. Therefore, the $2D$ or $3D$ nature of the HI determines the relation between $k$ and $\Lambda$.  
Assuming a cytoplasm viscosity $\eta_{3D,cyt} \in [10^{-1};10^{-2}]\, $Pa$\cdot$s~\cite{Margraves} and membrane
viscosity  $\eta_{2D,mem} \in [5\cdot10^{-7},10^{-8}]\,$ Pa$\cdot$s$\cdot$m~\cite{Cicuta}, we obtain $l=\eta_{2D,mem}/\eta_{3D,cyt} \in [10^{-1},10] \,
\mu$m that lies within the typical range of biological situations 
for which the typical velocity is $v \sim 0.1\, \mu m$~\cite{Campas-joanny}. The hopping rate can be assumed  $\omega_0\simeq 10^2 \alpha\,$s$^{-1}$,  $\alpha$ being the inverse of the efficiency~\footnote{$\omega_0$ has been calculated  assuming  instantaneous hopping  and motors  velocity  $v=1\mu$m/s. The time required to move one period $L\sim10\, $nm is $\tau=10^{-2}\, $s for $100 \%$ efficiency~\cite{Julicher1995}. For smaller efficiencies we estimate $\tau=1/\alpha10^{-2}\, $s.}.
For these values of parameters we get $\tau\simeq 0.2$ that fits in the range of values identified by Fig.~\ref{fig2}.(b).
By inverting Eq.~(\ref{k_2D}) we can calculate the dependence of $\Lambda$ upon $k$ when the $2D$ contribution dominates over the $3D$. Fig.~\ref{symmetry_breaking}.(a) shows that for $l \sim 1\mu$m, systems as small as $\Lambda \sim 0.5\, \mu$m can undergo hydrodynamically-induced symmetry breaking, while slightly larger systems $\Lambda \sim 0.7\, \mu$m  develop bistability. For larger values of $l$ the $3D$ contribution dominates and we expect hydrodynamically-induced symmetry breaking for systems of  order of $10-100\, \mu$m emphasizing the relevance of the $3D$ HI for larger systems such as neurons, or in technological applications as in microfluidic devices.
The onset of symmetry breaking and bistability depends on the parameters governing motors dynamics, namely, the force motor can provide, $f_0$ and their hopping rate $\bar\omega$. The ratio of $f_0$ and $\bar\omega$ is encoded in the dimensionless parameter, $\lambda$. As shown in Fig.~\ref{symmetry_breaking}.(b) the values of both $k_1$ and $k_2$ decrease upon increasing the strength of the motors or decreasing their hopping rate the latter being easily controlled in experiments by tuning the ATP concentration.

In conclusion, the HI between bidirectional molecular motors strongly affects their dynamics. For intermediate system sizes, HI triggers the onset of net motor currents~\footnote{Such a coupling is amplified for non processive motors that are more prone to show collective effects due to their low duty ratio~\cite{bloemink}.}. For these regimes the motor velocity has been observed to be oscillatory, about a non vanishing average, or bistable, when motors switch their direction over larger time scales, leading to an overall vanishing currents, as observed experimentally~\cite{Nebenfuhr01121999,Hendricks,Warshaw,goldman2010}. Such features rely on the local variation of the motor density, typical of HI, absent when this degree of freedom is neglected (rigid coupling) or reduced, as happens when the molecular motors are not conserved, e.g. through bath exchange. 
The typical system sizes over which the soft HI leads to symmetry breaking, or to bistability, are compatible with typical biologically relevant sizes, $\Lambda \in [0.1,10] \,\mu m$, typical for Golgi apparatus displacement~\cite{Nebenfuhr01121999} or bistable cargo transport~\cite{Hendricks,Warshaw,goldman2010}. 

P.M. acknowledges MINECO for
financial  support under  projects  FIS\  2015-67837-P and Institute Curie for financial support.
 I.P.  acknowledges MINECO and  DURSI for
financial  support under  projects  FIS\  2015-67837-P and  2014SGR-922,
respectively, and from  {\sl Generalitat de Catalunya  } under Program
{\sl Icrea Acad\`emia}.


\bibliography{joanny_letter,supl_mat}
\newpage
\begin{widetext}
\section{Discussion about the use of the Oseen Tensor for describing molecular motor hydrodynamic coupling}
In order to describe the hydrodynamic coupling between molecular motors pulling on a common cargo covered by a fluid-like membrane  we have exploited the Oseen tensor (see Eqs.(1),(2) in the main text). 
Indeed the Oseen tensor captures the far--field velocity profile in an unbound homogeneous fluid. However, in the present case molecular motors are running on a microtubule and the fluid flow they generate will be distorted by the presence of both the microtubule and the cargo motors are pulling on. 
Accordingly, in order to properly capture the presence of these boundaries, we should account for the appropriate image set. In some cases (e.g. for an infinite, solid planar wall) the presence of boundaries makes the leading contribution in the Green's function of the Stokes equation decay faster than the Stokeslet (captured by the Oseen tensor). However, even if the molecular motor has a diameter smaller than the cross section of a microtubule (even if not much smaller) the  screening induced by the microtubule will be weaker than the one corresponding to a planar, infinite wall. Accordingly, we expect that the details of the solid cylinder will modify the induced hydrodynamic field only quantitatively (and in hydrodynamics such changes are usually  gradual except very close to contact) with respect to the flow induced in an unbound system. 
Therefore, building the general framework on the Oseen tensor makes the derivations of the central equations and the analysis of their properties simpler and more transparent. 
Indeed, our approach allows us to identify the relevant coupling parameter, $k$. Even when the Oseen tensor is no longer relevant, the coupling parameter, $k$, remains the relevant parameter and only its functional dependence on the system geometry will be modified. Accordingly, our model is robust to the details of the hydrodynamic coupling and its results can be also meaningful in the context of coordinated molecular motor motion controlled by short range, unsteady  hydrodynamic coupling (described in Ref.[7]).


\section{Derivation of Eqs.(6) in the main text}

The dynamics of molecular motors is an intrinsic stochastic process. Therefore in order to gain analytical insight it is useful to study the ``typical'' behavior, i.e. averaged over the realization of the noise that affects the active stepping of the molecular motors. 
Following the approach developed in Ref.\cite{Guerin2011,GuerinPRE} we write down the functional Smoluchowsky equation governing the time evolution of the probability distribution $P[\tilde\rho,\tilde\sigma]$ of the dimensionless fields $\tilde{\rho}(x)$ and $\tilde{\sigma}(x)$:
\begin{eqnarray}~\label{eq:C-total_eq}
\dot{P}&=&\int dx\frac{\delta}{\delta\tilde{\rho}(x)}\left[\partial_{x}\left(\tilde{\rho}(x)v_\rho(x)\right)+\tilde\omega_{off}(x)\tilde{\rho}(x)-\tilde\omega_{on}(x)\tilde{\sigma}(x)\right]P+\nonumber\\
&&+\int dx\frac{\delta}{\delta\tilde{\sigma}(x)}\left[\partial_{x}\left(\tilde{\sigma}
(x)v_\sigma(x)\right)-\tilde\omega_{off}(x)\tilde{\rho}(x)+\tilde\omega_{on}
(x)\tilde\sigma(x)
+k_{\mbox{on}}\left( c_{\infty}-\rho(x)-\sigma(x)\right)-k_{\mbox{off}} \sigma(x)
\right]P
\end{eqnarray}
where $v_\rho(x)$ and $v_\sigma(x)$ are the local effective velocity field experienced by bound and weakly bound motors respectively and in the mean-field approach are defined as:
\begin{align}
v_\rho(x)&\simeq\frac{1}{\gamma}\left[f(x)+k\left\langle f(x)\left\langle\rho(x)\right\rangle_{\tilde\rho,\tilde\sigma}\right\rangle _{x}\right]\\
v_\sigma(x)&\simeq\frac{1}{\gamma}\left[k\left\langle f(x)\left\langle\rho(x)\right\rangle_{\tilde\rho,\tilde\sigma}\right\rangle _{x}\right]
\label{eq:vel_field1}
\end{align}
where with $\langle ... \rangle_{\tilde\rho,\tilde\sigma}$ we mean the average over all possible density distribution $\tilde\rho(x)$, $\tilde\sigma(x)$ each of which weighted with probability $P$ and with $\langle ... \rangle_{x}$ the average over all positions $x$, $x\in[0,L]$ weighted with constant probability $p(x)=dx/L$.  
From Eq.~\ref{eq:C-total_eq} we can write down the evolution equation for the average values of the fields  $\tilde{\rho}(x)$ and $\tilde{\sigma}(x)$:
\begin{eqnarray}
\left\langle \dot{\tilde\rho}(x)\right\rangle_{\tilde\rho,\tilde\sigma} &=&-\partial_{x}\lambda\left\langle \tilde\rho(x)\left(f(x)+k\left\langle f(x) \left\langle\tilde\rho(x)\right\rangle_{\tilde\rho,\tilde\sigma} \right\rangle _{x}\right)\right\rangle_{\tilde\rho,\tilde\sigma} -\tilde\omega_{off}(x)\left\langle \tilde\rho(x)\right\rangle_{\tilde\rho}+\tilde\omega_{on}(x)\left\langle \tilde\sigma(x)\right\rangle_{\tilde\sigma}\label{C-avg-equation-1}\\
\left\langle \dot{\tilde\sigma}(x)\right\rangle_{\tilde\rho,\tilde\sigma} &=&-\partial_{x}\lambda\left\langle \tilde\sigma(x)k\left\langle \left\langle f(x) \tilde\rho(x)\right\rangle_{\tilde\rho,\tilde\sigma}  \right\rangle_{x}\right\rangle_{\tilde\rho,\tilde\sigma} +\tilde\omega_{off}(x)\left\langle \tilde\rho(x)\right\rangle_{\tilde\rho}-\tilde\omega_{on}(x)\left\langle \tilde\sigma(x)\right\rangle_{\tilde\sigma}+\label{C-avg-equation-2}\\
&&+k_{\mbox{on}}\left( c_{\infty}-\langle\tilde{\rho}(x)\rangle_{\tilde{\rho}}-\langle\tilde{\sigma}(x)\rangle_{\tilde{\sigma}}\right)-k_{\mbox{off}} \langle\tilde{\sigma}(x)\rangle_{\tilde{\sigma}}\nonumber
\end{eqnarray}
By performing a mean--field approximation, namely assuming 
\begin{eqnarray}
 \left\langle \tilde\rho(x)\left\langle f(x) \left\langle\tilde\rho(x)\right\rangle_{\tilde\rho} \right\rangle _{x}\right\rangle_{\tilde\rho,\tilde\sigma}&=&\left\langle \tilde\rho(x)\right\rangle_{\tilde\rho,\tilde\sigma}\left\langle\left\langle f(x) \left\langle\tilde\rho(x)\right\rangle_{\tilde\rho,\tilde\sigma} \right\rangle _{x}\right\rangle_{\tilde\rho,\tilde\sigma}\\
 \left\langle \tilde\sigma(x)\left\langle f(x) \left\langle\tilde\rho(x)\right\rangle_{\tilde\rho} \right\rangle _{x}\right\rangle_{\tilde\rho,\tilde\sigma}&=&\left\langle \tilde\sigma(x)\right\rangle_{\tilde\rho,\tilde\sigma}\left\langle\left\langle f(x) \left\langle\tilde\rho(x)\right\rangle_{\tilde\rho,\tilde\sigma} \right\rangle _{x}\right\rangle_{\tilde\rho,\tilde\sigma}
\end{eqnarray}
and using the definition
\begin{eqnarray}
 \hat\rho(x)&=&\left\langle {\tilde\rho}(x)\right\rangle_{\tilde\rho,\tilde\sigma}\\
 \hat\sigma(x)&=&\left\langle {\tilde\sigma}(x)\right\rangle_{\tilde\rho,\tilde\sigma}
\end{eqnarray}
we can rewrite the previous equations as:
\begin{eqnarray}
\dot{\hat\rho}(x) &=&-\partial_{x}\lambda \hat\rho(x)\left[\tilde{f}(x)+k\left\langle \tilde{f}(x) \hat\rho(x)\right\rangle _{x}\right] -\tilde\omega_{off}(x)\hat\rho(x)+\tilde\omega_{on}(x)\hat\sigma(x)\label{C-avg-equation-1-mf}\\
\dot{\hat\sigma}(x) &=&-\partial_{x}\lambda\hat\sigma(x)k\left\langle \tilde{f}(x) \hat\rho(x)\right\rangle_{x} +\tilde\omega_{off}(x)\hat\rho(x)-\tilde\omega_{on}(x) \hat\sigma(x)+k_{\mbox{on}}\left( c_{\infty}-\hat{\rho}(x)-\hat{\sigma}(x)\right)-k_{\mbox{off}} \hat{\sigma}(x)\label{C-avg-equation-2-mf}
\end{eqnarray}

\section{Derivation of Eqs.(8) in the main text}

For fast bulk kinetics $k_{\mbox{on,off}}\rightarrow\infty$ Eq.~(\ref{C-avg-equation-2-mf}) reads:
\begin{align}
 \hat\sigma(x)=&\frac{k_{\mbox{on}}}{k_{\mbox{off}}+k_{\mbox{on}}}\left(c_\infty-\hat\rho(x)\right)
\end{align} 
substituting the last expression in the first of Eqs.(6) of the main text we obtain Eq.(8) of the main text.

\section{Derivation of Eqs.(9) in the main text}

The bifurcation portrait shown in Fig.1 of the main text has been obtained by numerically solving Eq.(8) (via a Lax-Wendroff method~\cite{Numerical-Recipes}) with periodic boundary conditions applied at the end of the extrema of the ratchet potential. In the following we derive an approximated scheme that allow us to discuss, from an analytical perspective, the bifurcations obtained numerically.
We start expressing $\hat\rho$ by its Fourier series:
\begin{equation}
 \hat\rho(x)=\rho_0+\Sigma_n \rho_n \cos(2\pi n x/L)+\bar\rho_n \sin(2\pi n x/L)
 \label{eq:supp-four-series}
\end{equation}
provides insight into the spontaneous symmetry breaking shown in Fig.1 of the main text. 
By substituting Eq.~(\ref{eq:supp-four-series}) into Eq.(8) of the main text and using Eq.(7) of the main text we obtain the following cascade of equations:
\begin{subequations}\label{eq:fourier}
\begin{align}
 \dot{\rho}_0&=\frac{1}{2}c_\infty\label{eq-1}\\
 \dot{\rho}_1&=-\frac{1}{2}\pi \lambda k \bar{\rho}_1\rho_1-\frac{1}{2} \pi \lambda \bar\rho_2 -\rho_1\\
 \dot{\bar\rho}_1&=\pi\lambda \rho_0 +\frac{1}{2}\pi \lambda k \rho^2_1+\frac{1}{2} \pi \lambda \rho_2 -\bar\rho_1-\frac{1}{2}\frac{\delta\omega}{\Delta\omega}c_\infty \\
 \dot{\rho}_n&=-\frac{1}{2}n\pi \lambda k \bar{\rho}_n\rho_1-\frac{1}{2} n\pi \lambda (\bar\rho_{n+1}+\bar\rho_{n-1}) -\rho_n\\
\dot{\bar\rho}_1&=\frac{1}{2}n\pi \lambda k \rho_1\rho_n+\frac{1}{2} n \pi \lambda (\rho_{n+1}+\rho_{n-1}) -\bar\rho_n\label{eq-2}
\end{align}
\end{subequations}
The set of Eqs.~(\ref{eq-1})-(\ref{eq-2}) is an infinite cascade that, in order to be treated analytically should be truncated. As a closure of the cascade we assume that the Fourier modes higher than the first ones (namely $\rho_1$ and $\bar \rho_1$) are decaying fast enough so that their amplitudes can be regarded as constants in the time scale in which $\rho_1$ and $\bar \rho_1$ are evolving. Accordingly, we regard $\rho_2$ and $\bar\rho_2$ as parameters whose values are obtained from the numerical solution of Eq.(8) and we get a closed system for $\rho_{1},\bar{\rho}_{1}$. Such a closure leads to good a match with the full numerical solutions.
In particular, such a system has a zero-velocity solution: 
\begin{eqnarray}
\rho_{1}&=&0,\bar{\rho}_{2}=0\\
\bar{\rho}_{1}&=&\frac{1}{\omega_{0}}\left(\pi\lambda\rho_{0}+\frac{1}{2}\pi\lambda\rho_{2}-\frac{1}{4}c_\infty\right)
\end{eqnarray}
and, for $\rho_{1}\neq0$ a, possibly, moving solution characterized by 
\begin{eqnarray}
\bar{\rho}_{1}=\frac{-2}{\pi\lambda k\rho_{1}}\left(\frac{1}{2}\pi\lambda\bar{\rho}_{2}+\rho_{1}\right)
\end{eqnarray}
where $\rho_{1}$ is obtained by solving:
\begin{equation}
2\left(\pi\lambda k\right)^{2}\rho_{1}^{3}+\left(4k\left(\pi\lambda\right)^{2}\left(\rho_{0}+\frac{1}{2}\rho_{2}\right)+8-2\frac{\delta\omega}{\Delta\omega}\pi\lambda k c_\infty\right)\rho_{1}+4\pi\lambda \bar{\rho}_{2}=0
\label{eq:solutions}
\end{equation}
The necessary condition in order to observe a spontaneous symmetry breaking is that Eq.~(\ref{eq:solutions}) must have three real solutions. For the case $\bar\rho_2=0$ (that coincides with the outcome of our numerical solutions) we can write down a necessary condition for having three real solutions of Eq.~(\ref{eq:solutions}), namely:
\begin{equation}
4\pi^{2}k\lambda^{2}\left(\rho_{0}+\frac{1}{2}\rho_{2}\right)+8-2\frac{\delta\omega}{\Delta\omega}k\pi\lambda c_\infty<0.
\label{eq:sol_condition}
\end{equation}
For rigidly coupled motors the first term in Eq.~(\ref{eq:sol_condition}) vanishes and Eq.~(\ref{eq:sol_condition}) is a necessary and sufficient condition~\cite{Guerin2011} for the onset of a moving solution. In this case, by rearranging eq.~\ref{eq:sol_condition}, the necessary and sufficient condition for symmetry breaking reads: 
\begin{equation}
k\lambda\ge\frac{8}{\pi c_\infty}\,.
\label{necess-cond-gamma-0}
\end{equation}
According to Eq.~(\ref{necess-cond-gamma-0}) by increasing the dimensionless forcing $\lambda$ we can diminish the minimum value of the coupling above which  symmetry breaking occurs. For soft-hydrodynamic coupled motors the necessary, but not sufficient, condition for symmetry breaking is  
\begin{equation}
k\geq\frac{8}{\pi\lambda\left(2\frac{\delta\omega}{\Delta\omega}c_\infty-2\pi\lambda(2\rho_0+\rho_2)\right)}.
\label{necess-cond-gamma-1}
\end{equation}
For small values of $\rho_2$, as obtained from the numerical solutions of Eq.(8), we can assume  $\rho_2\simeq 0$ and, using Eq.(\ref{eq-1}), the last expression simplifies to
\begin{equation}
k\geq\frac{4}{\pi\lambda\left(\frac{\delta\omega}{\Delta\omega}-\pi\lambda\right)c_\infty}.
\label{eq:k_c}
\end{equation}
i.e. Eq.(9) of the main text.
Since $k\ge0$, Eq.~(\ref{eq:k_c}) identifies an interval $\lambda \epsilon [0;\lambda_{max}]$ for which symmetry breaking occurs, with $\lambda_{max}= \frac{1}{\pi}\frac{\delta \omega}{\Delta \omega}$.
According to Eq.~(\ref{eq:k_c}), $k$ diverges for both $\lambda=0,\lambda=\lambda_{max}$ leading to the existence of a minimum value $k=k_c$ at $\lambda=\lambda_c=\frac{1}{2}\lambda_{max}$.
Therefore Eq.~(\ref{eq:k_c}) provides a \textit{prediction} of the critical value, $\lambda_c$ above which we can observe spontaneous symmetry breaking. Substituting the values of $\delta \omega/\Delta \omega$ used in the numerical solutions of Eq.(8) that is, the full set of Eqs.(\ref{eq:fourier}), into Eq.~(\ref{eq:k_c}) we obtain $\lambda_c\simeq 8\cdot 10^{-2}$ that is in good agreement with the value obtained from the full solution of Eq.(8) shown in Fig.1. Such an agreement between the numerical solutions and the approximated analytical stability analysis underlines the relevance of the slowest modes in the control of the instability. We remark that the linear stability analysis performed here is general and it is valid for all models that share the same functional form once linearized about the non--motile state. 
Therefore, since such a functional will not be affected by accounting for more detailed models for the dynamics of the molecular motors or by solving the Stokes equation beyond the Oseen approximation, the agreement between the linear stability analysis and the numerical solution shows that the phenomenology captured by our model is robust and will persist also for more detailed models that go beyond the mean--field approximations approach. 
\newpage
\section{Bistable regime}

As explained in the main text, Fig.1 shows that velocity oscillates with frequency $\omega$ about a non vanishing value whose sign switches with frequency $\Omega$.
\begin{figure}[h!]
\centering
 \includegraphics[scale=0.5]{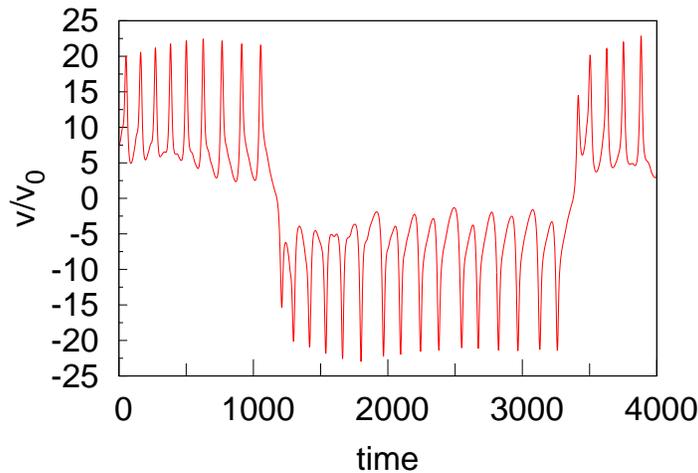}
 \caption{Time dependence of the velocity of bound motors in the bistable regime.}
\end{figure}

\bibliography{supl_mat}

\begin{thebibliography}{22}
\expandafter\ifx\csname natexlab\endcsname\relax\def\natexlab#1{#1}\fi
\expandafter\ifx\csname bibnamefont\endcsname\relax
  \def\bibnamefont#1{#1}\fi
\expandafter\ifx\csname bibfnamefont\endcsname\relax
  \def\bibfnamefont#1{#1}\fi
\expandafter\ifx\csname citenamefont\endcsname\relax
  \def\citenamefont#1{#1}\fi
\expandafter\ifx\csname url\endcsname\relax
  \def\url#1{\texttt{#1}}\fi
\expandafter\ifx\csname urlprefix\endcsname\relax\def\urlprefix{URL }\fi
\providecommand{\bibinfo}[2]{#2}
\providecommand{\eprint}[2][]{\url{#2}}

\bibitem[{\citenamefont{Howard}(2001)}]{Howard}
\bibinfo{author}{\bibfnamefont{J.}~\bibnamefont{Howard}},
  \emph{\bibinfo{title}{Mechanics of Motor Proteins and Cytoskeleton}}
  (\bibinfo{publisher}{Sinauer}, \bibinfo{address}{Sunderland},
  \bibinfo{year}{2001}).

\bibitem[{\citenamefont{Malgaretti et~al.}(2012)\citenamefont{Malgaretti,
  Pagonabarraga, and Frenkel}}]{Malgaretti}
\bibinfo{author}{\bibfnamefont{P.}~\bibnamefont{Malgaretti}},
  \bibinfo{author}{\bibfnamefont{I.}~\bibnamefont{Pagonabarraga}},
  \bibnamefont{and} \bibinfo{author}{\bibfnamefont{D.}~\bibnamefont{Frenkel}},
  \bibinfo{journal}{Phys. Rev. Lett.} \textbf{\bibinfo{volume}{109}},
  \bibinfo{pages}{168101} (\bibinfo{year}{2012}).

\bibitem[{\citenamefont{Nelson et~al.}(2014)\citenamefont{Nelson, Trybus, and
  Warshaw}}]{Nelson2014}
\bibinfo{author}{\bibfnamefont{S.}~\bibnamefont{Nelson}},
  \bibinfo{author}{\bibfnamefont{K.}~\bibnamefont{Trybus}}, \bibnamefont{and}
  \bibinfo{author}{\bibfnamefont{D.}~\bibnamefont{Warshaw}},
  \bibinfo{journal}{Proc. Nat. Acad. Sci.} \textbf{\bibinfo{volume}{111}},
  \bibinfo{pages}{E3986} (\bibinfo{year}{2014}).

\bibitem[{\citenamefont{Ma and Chisholm}(2002)}]{Ma}
\bibinfo{author}{\bibfnamefont{S.}~\bibnamefont{Ma}} \bibnamefont{and}
  \bibinfo{author}{\bibfnamefont{R.~L.} \bibnamefont{Chisholm}},
  \bibinfo{journal}{J. Cell. Sci.} \textbf{\bibinfo{volume}{115}},
  \bibinfo{pages}{1453} (\bibinfo{year}{2002}).

\bibitem[{\citenamefont{Yusuf~Ali et~al.}(2011)\citenamefont{Yusuf~Ali,
  Kennedy, Safer, Trybus, Lee~Sweeney, and Warshaw}}]{Warshaw}
\bibinfo{author}{\bibfnamefont{M.}~\bibnamefont{Yusuf~Ali}},
  \bibinfo{author}{\bibfnamefont{G.~G.} \bibnamefont{Kennedy}},
  \bibinfo{author}{\bibfnamefont{D.}~\bibnamefont{Safer}},
  \bibinfo{author}{\bibfnamefont{M.}~\bibnamefont{Trybus}},
  \bibinfo{author}{\bibfnamefont{H.}~\bibnamefont{Lee~Sweeney}},
  \bibnamefont{and} \bibinfo{author}{\bibfnamefont{D.~M.}
  \bibnamefont{Warshaw}}, \bibinfo{journal}{Proc. Nat. Acad. Sci.}
  \textbf{\bibinfo{volume}{108}}, \bibinfo{pages}{535} (\bibinfo{year}{2011}).

\bibitem[{\citenamefont{Kunwar et~al.}(2011)\citenamefont{Kunwar, Tripathy, Xu,
  Mattson, Anand, Sigua, Vershinin, McKenney, Yu, Mogliner
  et~al.}}]{Kunwar2011}
\bibinfo{author}{\bibfnamefont{A.}~\bibnamefont{Kunwar}},
  \bibinfo{author}{\bibfnamefont{S.~K.} \bibnamefont{Tripathy}},
  \bibinfo{author}{\bibfnamefont{J.}~\bibnamefont{Xu}},
  \bibinfo{author}{\bibfnamefont{M.~K.} \bibnamefont{Mattson}},
  \bibinfo{author}{\bibfnamefont{P.}~\bibnamefont{Anand}},
  \bibinfo{author}{\bibfnamefont{R.}~\bibnamefont{Sigua}},
  \bibinfo{author}{\bibfnamefont{M.}~\bibnamefont{Vershinin}},
  \bibinfo{author}{\bibfnamefont{R.~J.} \bibnamefont{McKenney}},
  \bibinfo{author}{\bibfnamefont{C.~C.} \bibnamefont{Yu}},
  \bibinfo{author}{\bibfnamefont{A.}~\bibnamefont{Mogliner}},
  \bibnamefont{et~al.}, \bibinfo{journal}{Proc. Nat. Acad. Sci.}
  \textbf{\bibinfo{volume}{108}}, \bibinfo{pages}{18960}
  (\bibinfo{year}{2011}).

\bibitem[{\citenamefont{Badoual et~al.}(2002)\citenamefont{Badoual,
  J\"{u}licher, and Prost}}]{badoual}
\bibinfo{author}{\bibfnamefont{M.}~\bibnamefont{Badoual}},
  \bibinfo{author}{\bibfnamefont{F.}~\bibnamefont{J\"{u}licher}},
  \bibnamefont{and} \bibinfo{author}{\bibfnamefont{J.}~\bibnamefont{Prost}},
  \bibinfo{journal}{Proc. Nat. Acad. Sci.} \textbf{\bibinfo{volume}{99}},
  \bibinfo{pages}{6696} (\bibinfo{year}{2002}).

\bibitem[{\citenamefont{Nebenf\"uhr et~al.}(1999)\citenamefont{Nebenf\"uhr,
  Gallagher, Dunahay, Frohlick, Mazurkiewicz, Meehl, and
  Staehelin}}]{Nebenfuhr01121999}
\bibinfo{author}{\bibfnamefont{A.}~\bibnamefont{Nebenf\"uhr}},
  \bibinfo{author}{\bibfnamefont{L.~A.} \bibnamefont{Gallagher}},
  \bibinfo{author}{\bibfnamefont{T.~G.} \bibnamefont{Dunahay}},
  \bibinfo{author}{\bibfnamefont{J.~A.} \bibnamefont{Frohlick}},
  \bibinfo{author}{\bibfnamefont{A.~M.} \bibnamefont{Mazurkiewicz}},
  \bibinfo{author}{\bibfnamefont{J.~B.} \bibnamefont{Meehl}}, \bibnamefont{and}
  \bibinfo{author}{\bibfnamefont{L.~A.} \bibnamefont{Staehelin}},
  \bibinfo{journal}{Plant Physiology} \textbf{\bibinfo{volume}{121}},
  \bibinfo{pages}{1127} (\bibinfo{year}{1999}).

\bibitem[{\citenamefont{Hendricks et~al.}(2010)\citenamefont{Hendricks,
  Perlson, Ross, Schroeder~III, Tokito, and Holzbaur}}]{Hendricks}
\bibinfo{author}{\bibfnamefont{A.}~\bibnamefont{Hendricks}},
  \bibinfo{author}{\bibfnamefont{E.}~\bibnamefont{Perlson}},
  \bibinfo{author}{\bibfnamefont{J.~L.} \bibnamefont{Ross}},
  \bibinfo{author}{\bibfnamefont{H.~W.} \bibnamefont{Schroeder~III}},
  \bibinfo{author}{\bibfnamefont{M.}~\bibnamefont{Tokito}}, \bibnamefont{and}
  \bibinfo{author}{\bibfnamefont{E.}~\bibnamefont{Holzbaur}},
  \bibinfo{journal}{Curr. Biol.} \textbf{\bibinfo{volume}{20}},
  \bibinfo{pages}{697} (\bibinfo{year}{2010}).

\bibitem[{\citenamefont{Goldman and Holzbaur}(2010)}]{goldman2010}
\bibinfo{author}{\bibfnamefont{Y.}~\bibnamefont{Goldman}} \bibnamefont{and}
  \bibinfo{author}{\bibfnamefont{E.}~\bibnamefont{Holzbaur}},
  \bibinfo{journal}{Curr. Op. Cell. Biol.} \textbf{\bibinfo{volume}{22}},
  \bibinfo{pages}{4} (\bibinfo{year}{2010}).

\bibitem[{\citenamefont{J\"ulicher and Prost}(1995)}]{Julicher1995}
\bibinfo{author}{\bibfnamefont{F.}~\bibnamefont{J\"ulicher}} \bibnamefont{and}
  \bibinfo{author}{\bibfnamefont{J.}~\bibnamefont{Prost}},
  \bibinfo{journal}{Phys. Rev. Lett.} \textbf{\bibinfo{volume}{75}},
  \bibinfo{pages}{2618} (\bibinfo{year}{1995}).

\bibitem[{\citenamefont{J\"ulicher and Prost}(1997)}]{Julicher1997}
\bibinfo{author}{\bibfnamefont{F.}~\bibnamefont{J\"ulicher}} \bibnamefont{and}
  \bibinfo{author}{\bibfnamefont{J.}~\bibnamefont{Prost}},
  \bibinfo{journal}{Phys. Rev. Lett.} \textbf{\bibinfo{volume}{78}},
  \bibinfo{pages}{4510} (\bibinfo{year}{1997}).

\bibitem[{\citenamefont{Klumpp et~al.}(2010)\citenamefont{Klumpp, Lipowsky, and
  Mu}}]{Lipowsky2010}
\bibinfo{author}{\bibfnamefont{S.}~\bibnamefont{Klumpp}},
  \bibinfo{author}{\bibfnamefont{R.}~\bibnamefont{Lipowsky}}, \bibnamefont{and}
  \bibinfo{author}{\bibfnamefont{M.~J.~I.} \bibnamefont{Mu}},
  \bibinfo{journal}{Biophysical Journal} \textbf{\bibinfo{volume}{98}},
  \bibinfo{pages}{2610} (\bibinfo{year}{2010}).

\bibitem[{\citenamefont{Gu\'erin
  et~al.}(2011{\natexlab{a}})\citenamefont{Gu\'erin, Prost, and
  Joanny}}]{Guerin2011}
\bibinfo{author}{\bibfnamefont{T.}~\bibnamefont{Gu\'erin}},
  \bibinfo{author}{\bibfnamefont{J.}~\bibnamefont{Prost}}, \bibnamefont{and}
  \bibinfo{author}{\bibfnamefont{J.-F.} \bibnamefont{Joanny}},
  \bibinfo{journal}{Phys. Rev. Lett.} \textbf{\bibinfo{volume}{106}},
  \bibinfo{pages}{068101} (\bibinfo{year}{2011}{\natexlab{a}}).

\bibitem[{\citenamefont{Aghababaie et~al.}(1999)\citenamefont{Aghababaie,
  Menon, and Plischke}}]{Aghababaie}
\bibinfo{author}{\bibfnamefont{Y.}~\bibnamefont{Aghababaie}},
  \bibinfo{author}{\bibfnamefont{G.~I.} \bibnamefont{Menon}}, \bibnamefont{and}
  \bibinfo{author}{\bibfnamefont{M.}~\bibnamefont{Plischke}},
  \bibinfo{journal}{Phys. Rev. E} \textbf{\bibinfo{volume}{59}},
  \bibinfo{pages}{2578} (\bibinfo{year}{1999}).

\bibitem[{\citenamefont{Gu\'erin et~al.}(2010)\citenamefont{Gu\'erin, Prost,
  Martin, and Joanny}}]{Guerin-review}
\bibinfo{author}{\bibfnamefont{T.}~\bibnamefont{Gu\'erin}},
  \bibinfo{author}{\bibfnamefont{J.}~\bibnamefont{Prost}},
  \bibinfo{author}{\bibfnamefont{P.}~\bibnamefont{Martin}}, \bibnamefont{and}
  \bibinfo{author}{\bibfnamefont{J.-F.} \bibnamefont{Joanny}},
  \bibinfo{journal}{Curr. Op. Cell Biol.} \textbf{\bibinfo{volume}{22}},
  \bibinfo{pages}{14} (\bibinfo{year}{2010}).

\bibitem[{\citenamefont{Margraves et~al.}(2011)\citenamefont{Margraves, Kihm,
  Yoon, Choi, Liggett, and Baek}}]{Margraves}
\bibinfo{author}{\bibfnamefont{C.}~\bibnamefont{Margraves}},
  \bibinfo{author}{\bibfnamefont{K.}~\bibnamefont{Kihm}},
  \bibinfo{author}{\bibfnamefont{S.~Y.} \bibnamefont{Yoon}},
  \bibinfo{author}{\bibfnamefont{S.}~\bibnamefont{Choi},
  \bibfnamefont{C.~K.~Lee}},
  \bibinfo{author}{\bibfnamefont{J.}~\bibnamefont{Liggett}}, \bibnamefont{and}
  \bibinfo{author}{\bibfnamefont{S.~J.} \bibnamefont{Baek}},
  \bibinfo{journal}{Biotechnology and Bioengineering}
  \textbf{\bibinfo{volume}{108}}, \bibinfo{pages}{2504} (\bibinfo{year}{2011}).

\bibitem[{\citenamefont{Cicuta et~al.}(2007)\citenamefont{Cicuta, Keller, and
  Veatch}}]{Cicuta}
\bibinfo{author}{\bibfnamefont{P.}~\bibnamefont{Cicuta}},
  \bibinfo{author}{\bibfnamefont{S.~L.} \bibnamefont{Keller}},
  \bibnamefont{and} \bibinfo{author}{\bibfnamefont{S.~L.}
  \bibnamefont{Veatch}}, \bibinfo{journal}{J. Phys. Chem. B}
  \textbf{\bibinfo{volume}{111}}, \bibinfo{pages}{3328} (\bibinfo{year}{2007}).

\bibitem[{\citenamefont{Campas et~al.}(2008)\citenamefont{Campas, Leduc,
  Bassereau, Casademunt, Joanny, and Prost}}]{Campas-joanny}
\bibinfo{author}{\bibfnamefont{O.}~\bibnamefont{Campas}},
  \bibinfo{author}{\bibfnamefont{C.}~\bibnamefont{Leduc}},
  \bibinfo{author}{\bibfnamefont{P.}~\bibnamefont{Bassereau}},
  \bibinfo{author}{\bibfnamefont{J.}~\bibnamefont{Casademunt}},
  \bibinfo{author}{\bibfnamefont{J.-F.} \bibnamefont{Joanny}},
  \bibnamefont{and} \bibinfo{author}{\bibfnamefont{J.}~\bibnamefont{Prost}},
  \bibinfo{journal}{Biophysical J.} \textbf{\bibinfo{volume}{94}},
  \bibinfo{pages}{5009} (\bibinfo{year}{2008}).

\bibitem[{\citenamefont{Bloemink and Geeves}(2011)}]{bloemink}
\bibinfo{author}{\bibfnamefont{M.}~\bibnamefont{Bloemink}} \bibnamefont{and}
  \bibinfo{author}{\bibfnamefont{M.}~\bibnamefont{Geeves}},
  \bibinfo{journal}{Seminars in Cell and Developmental Biology}
  \textbf{\bibinfo{volume}{22}}, \bibinfo{pages}{961} (\bibinfo{year}{2011}).

\bibitem[{\citenamefont{Gu\'erin
  et~al.}(2011{\natexlab{b}})\citenamefont{Gu\'erin, Prost, and
  Joanny}}]{GuerinPRE}
\bibinfo{author}{\bibfnamefont{T.}~\bibnamefont{Gu\'erin}},
  \bibinfo{author}{\bibfnamefont{J.}~\bibnamefont{Prost}}, \bibnamefont{and}
  \bibinfo{author}{\bibfnamefont{J.-F.} \bibnamefont{Joanny}},
  \bibinfo{journal}{Phys. Rev. E} \textbf{\bibinfo{volume}{87}},
  \bibinfo{pages}{032601} (\bibinfo{year}{2011}{\natexlab{b}}).

\bibitem[{\citenamefont{Press et~al.}(2007)\citenamefont{Press, Teukolsky,
  Vetterling, and Flannery}}]{Numerical-Recipes}
\bibinfo{author}{\bibfnamefont{W.~H.} \bibnamefont{Press}},
  \bibinfo{author}{\bibfnamefont{S.~A.} \bibnamefont{Teukolsky}},
  \bibinfo{author}{\bibfnamefont{W.~T.} \bibnamefont{Vetterling}},
  \bibnamefont{and} \bibinfo{author}{\bibfnamefont{B.~P.}
  \bibnamefont{Flannery}}, \emph{\bibinfo{title}{Numerical Recipes, third
  edition}} (\bibinfo{publisher}{Cambridge University Press},
  \bibinfo{address}{Cambridge}, \bibinfo{year}{2007}).

\end{thebibliography}
\end{widetext}
\end{document}